\documentclass[showpacs,showkeys,pra]{revtex4}

\usepackage{amsmath}
\usepackage{amssymb}
\usepackage{graphicx}
\usepackage{color}

\begin{document}

\title{Self-maintaining defect/droplets from two interacting Bose-Einstein condensates}

\author{Vladimir Dzhunushaliev}
\email{vdzhunus@krsu.edu.kg}
\affiliation{
%Dept. Theor. and Nucl. Phys., KazNU, Almaty, 010008, Kazakhstan 
Institute for Basic Research,
Eurasian National University,
Astana, 010008, Kazakhstan
}

\date{\today}

\begin{abstract}
We consider two interacting Bose-Einstein condensates (BEC's) with different kind of the potential energy of interaction of the condensates: (a) the standard potential; (b) the potential has a positive three-body and a negative two-body scattering terms and (c) the potential has a positive four-body and a negative three-body scattering terms for the first BEC and a positive three-body and a negative two-body scattering terms for the second BEC. It is shown that in these cases there exist regular spherically symmetric solutions. Physically such solution is either a defect or a droplet created by the condensates. The defect is a cavity filled with one BEC on the background of another BEC. The droplet is an object on the background of the empty space. For (a) and (b) cases the obtained objects are supported by a constant external trapping potential and for (c) case the droplet is a self-maintaining object without any external potential. The possibility of construction of an elementary logic qubit device on the basis of this droplet is discussed. 
\end{abstract}


\maketitle

\section{Introduction}

Usually two interacting Bose-Einstein condensates are described by equations where the potential energy of interaction of the condensates has terms of 2-$th$ and 4-$th$ orders \cite{Ho}. But in Ref. \cite{Bulgac}  it was shown that the Hamiltonian of two interacting BEC's may have the interactions terms with both 6-$th$ and 4-$th$ orders. There was shown that the term of 6-$th$ order can be positive and the term of 4-$th$ order can be negative. Here we would like consider three different cases of two interacting BEC's: (a) both BEC's have standard interaction scattering terms and interact with a constant external potential; (b) both BEC's have positive three-body and negative two-body scattering terms in the potential and interact with a constant external potential; (c) one BEC is as in the previous item but another BEC has positive four-body and negative three-body scattering terms and they do not interact with any external potential. 

We will show that in such coupled equations set (describing two interacting BEC's) regular solutions appear. From our point of view such solutions describe either defect or a droplet created by two interacting BEC's. We will show that: in the case (a) we have a defect supported by a constant external potential; in (b) we have a droplet trapped by a constant external potential and in (c) we have a droplet without any support of an external potential. The defect means that we have a cavity on the background of the space filled with a first kind of BEC. The first kind of BEC has a constant energy density at the infinity. The cavity is filled with another BEC whose energy density asymptotically is zero. The droplet means that we have an object filled with two BEC's with both asymptotically zero energy densities. 

In Ref. \cite{Bulgac} it is shown that if three-body term is repulsive (the interactions terms of  6-$th$ order in our language) and two-body term is attractive (the interactions terms of  4-$th$ order in our language) then the conclusion is made: "Since the two - body contribution to the ground state energy of a dilute Bose gas is negative, the three - body collisions in the regime where $g_3 > 0$ could lead to the stabilization of the system. What is particularly interesting for such a system is that a boson droplet  a boselet  could become selfbound and the trapping potential is not required anymore to keep the particles together". Below we will show that such self-maintaining configuration (droplet or boselet) does exist as the solution of equations \eqref{3-30} \eqref{3-40} if one BEC has a positive four-body and negative three-body scattering terms and another BEC has a positive three-body and negative two-body scattering terms.

In this work we investigate the possibility of the existence of soliton - like solutions for two interacting Bose - Einstein condensates \textcolor{blue}{\emph{with and without any external trapping potential}}. The soliton-like solutions on the background of a trapping potential are widely discussed. In Ref. \cite{Kumar} the coupled  Gross - Pitaevskii equations describing the dynamics of two hyperfine states 
of Bose - Einstein condensates are investigated and the integrability condition for the propagation of bright vector solitons is deduced. In Ref. \cite{QiuYan} the authors investigate the combined soliton solutions of two - component Bose - Einstein condensates with external potential. The results show that the intraspecies (interspecies) interaction strengths and the external trapped potential clearly affect the formation of darkdark, brightbright, and darkbright soliton solutions in different regions. In Ref. \cite{Deng} the localized nonlinear matter waves in the two - component Bose - Einstein condensates with time- and space-modulated nonlinearities  analytically  and  numerically is investigated. The dynamics of these matter waves, including the breathing solitons, quasibreathing solitons, resonant solitons, and moving solitons, is discussed. In Ref. \cite{Baizakov} sound waves in two - component Bose - Einstein condensates is investigated and  proposed a new method of wave generation which is based on a fast change of the inter - species interaction constant. In Ref. \cite{Kawaja} stable skyrmions in two component Bose - Einstein condensates are considered. In Ref. \cite{Law} the authors report the numerical realization of robust 2 - component structures in 2d and 3d BEC's with non - trivial topological charge in one component. The vortex - bright solitary waves are found to be very robust in both in the homogeneous medium and in the presence of parabolic and periodic external confinement. In Ref. \cite{Zhang} a family of exact vector - soliton solutions for the coupled nonlinear Schr\"odinger equations with tunable interactions and harmonic potential is presented. 

\section{Defect solution with two-body scattering terms and external trapping constant potential}
\label{twoBody}

In this section we consider a defect solution. The solution describes the defect filled with one kind of BEC and placed in the space filled with another kind of BEC. 

\begin{figure}[h]
\begin{minipage}[t]{.45\linewidth}
  \begin{center}
  \fbox{
  \includegraphics[width=1.\textwidth]{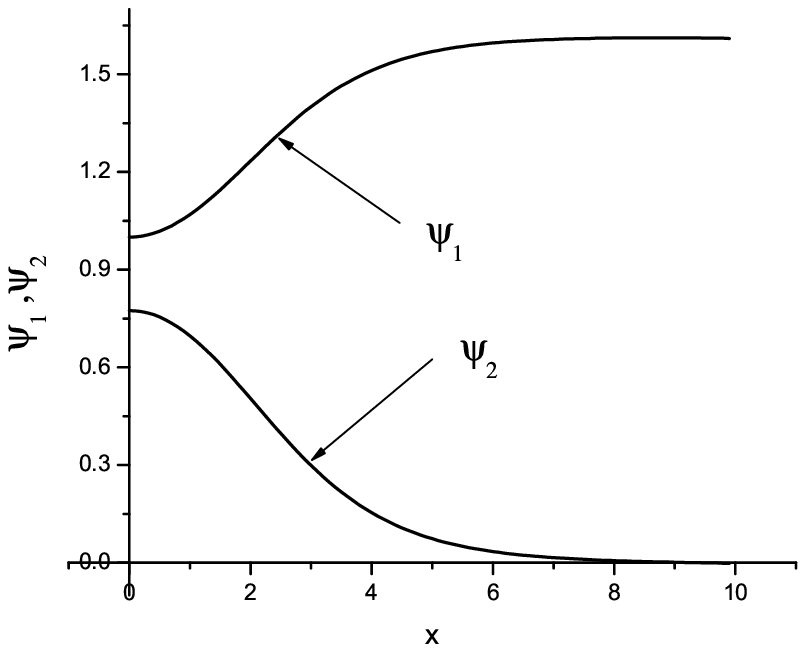}}
  \caption{The profiles of $\psi_{1,2}(x)$, $\mu_1 = 1.61716$, 
  $\mu_2 = 1.49276$}
  \label{sec1fig1}
  \end{center}
\end{minipage}\hfill
\begin{minipage}[t]{.45\linewidth}
  \begin{center}
  \fbox{
		\includegraphics[width=1.\textwidth]{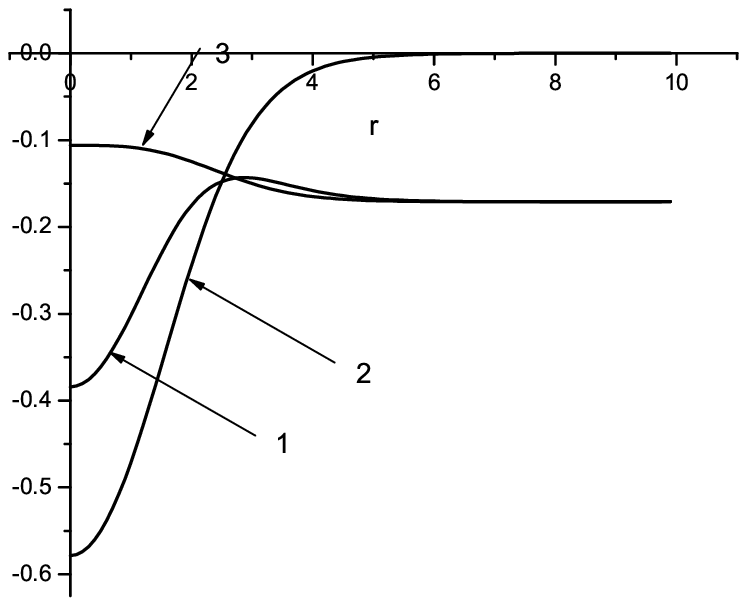}}
	  \caption{The profiles the dimensionless energy densities: 
	  curve 1 - the dimensionless energy density \eqref{1-130}; 
	  curve 2 - the dimensionless energy density 
	  of the BEC with $\psi_2$, \eqref{1-134};
	  curve 3 - the dimensionless energy density 
	  of the BEC with $\psi_2$, \eqref{1-138}.}
  \label{sec1fig2}
  \end{center}
\end{minipage}\hfill 
\end{figure}

Considering a two-component BEC, the behavior of the condensates that are prepared in two hyperfine states can be described at sufficiently low temperatures by the two-coupled GP equation of the following form \cite{Bulgac} 
\begin{eqnarray}
	i \hbar \frac{\partial \tilde \psi_1}{\partial t} &=& 
	\left(
		- \frac{\hbar^2}{2m_1} \nabla^2 + U_{11} \left| \tilde \psi_1 \right|^2 + 
		U_{12} \left| \tilde \psi_2 \right|^2 + V_{1} 
	\right) \tilde \psi_1 , 
\label{1-10} \\
	i \hbar \frac{\partial \tilde \psi_2}{\partial t} &=& 
	\left(
		- \frac{\hbar^2}{2m_2} \nabla^2 + U_{22} \left| \tilde \psi_2 \right|^4 + 
		U_{21} \left| \tilde \psi_1 \right|^2 + V_2 
	\right) \tilde \psi_2 
\label{1-20}
\end{eqnarray}
where  the  condensate  wave  functions  are  normalized  by  particle  numbers $N_i = \int \left| \tilde \psi_i \right|^2 dV$; $V_{1,2}$ are  external trapping potentials. The resulting equations for the wave functions $\psi_{1,2}(\vec r,t)$ in dimensionless form can be written as
\begin{eqnarray}
	i \frac{\partial \psi_1}{\partial t} &=& 
	\left(
		- \nabla^2 + u_{11} \left| \psi_1 \right|^2 + u_{12} \left| \psi_2 \right|^2 
		+ v_1 
	\right) \psi_1 , 
\label{1-30} \\
	i \frac{\partial \psi_2}{\partial t} &=& 
	\left(
		- k \nabla^2 + u_{22} \left| \psi_2 \right|^2 + u_{21} \left| \psi_1 \right|^2 
		+ v_2 
	\right) \psi_2 
\label{1-40}
\end{eqnarray}
here $k=m_1/m_2$ and we redefined $t/t_0 \rightarrow t$; $\vec r/l_0 \rightarrow \vec r$; $t_0=\frac{2m_1}{\hbar \psi_1^{2/3}(0)}$; $l_0 = \psi_1^{-1/3}(0)$; $\psi_{1,2}=\frac{\tilde \psi_{1,2}}{\psi_1(0)}$; $u_{ii}=\frac{2m_1 U_{ii} \psi_1^{4/3}(0)}{\hbar^2}$;  $u_{12}=\frac{2m_1 U_{12} \psi_1^{4/3}(0)}{\hbar^2}$ and $v_i = \frac{2 m_1 V_i}{\hbar^2 \psi_1^{2/3}(0)}$. For the simplicity we will  consider the case $k=1$. We are searching for a static  spherical symmetric solution: $\psi_{1,2}(\vec r,t) = \psi_{1,2}(r)$. In this case Eq's \eqref{1-30} \eqref{1-40} are 
\begin{eqnarray}
	\psi_1'' + \frac{2}{r} \psi_1' &=& \psi_1 \left[
		\lambda_3 \psi_2^2 + \lambda_1 \left(  
			\psi_1^2 - \mu_1^2
		\right)
	\right],
\label{1-50} \\
	\psi_2'' + \frac{2}{r} \psi_2' &=& \psi_2 \left[
		\lambda_3 \psi_1^2 + \lambda_2 \left(  
			\psi_2^2 - \mu_2^2
		\right)
	\right].
\label{1-60}
\end{eqnarray}
where $\lambda_{i} = u_{ii}, i=1,2$; $\lambda_{i} \mu_{i}^2 = - v_{i}$; $\lambda_3=u_{12}=u_{21}$. Eq's \eqref{1-50} \eqref{1-60} are written in the form which is convenient for the numerical calculations. For the numerical calculations we choose the boundary conditions and parameters values as follows 
\begin{align}
	\lambda_1 &=0.1; &&\lambda_2 = 1. ; &&\lambda_3 = 1. ; 
\label{1-70} \\
	\psi_1(0) &=1; &&\psi_1'(0) = 0; &&
\label{1-80} \\
	\psi_2(0) &=\sqrt{0.6};  &&\psi_2'(0) = 0 . &&
\label{1-90}
\end{align}
The solution is searching as the nonlinear eigenvalue problem: $\mu_{1,2}$ are eigenvalues and corresponding functions $\psi_{1,2}$ are eigenfunctions. We solve equations set \eqref{1-50} \eqref{1-60} numerically. The profiles of functions $\psi_{1,2}$ in Fig. \ref{sec1fig1} are presented. 

The asymptotic behavior of the functions $\psi_{1,2}$ is following 
\begin{eqnarray}
	\psi_1 & \approx & \mu_1 - \psi_{1, \infty} 
	\frac{e^{-r \sqrt{2 \lambda_1 \mu_1^2}}}{r} , 
\label{1-110} \\
	\psi_2 & \approx & \psi_{2, \infty} 
	\frac{e^{-r \sqrt{\mu_1^2 - \lambda_2 \mu_2^2}}}{r}
\label{1-120} 
\end{eqnarray}
where $\psi_{1,2, \infty}$ are constants. 

One can consider Eq's \eqref{1-50} \eqref{1-60} as Euler - Lagrangian equations. Then the dimensionless energy density has the form 
\begin{eqnarray}
	\varepsilon(\psi_{1,2}) &=& 
	\frac{1}{2} \left( \nabla \psi_1 \right)^2 + 
	\frac{1}{2} \left ( \nabla \psi_2 \right )^2 + 
	V\left( \psi_{1,2} \right) ,
\label{1-130}\\
	V\left( \psi_{1,2} \right) &=& 
		\frac{\lambda_1}{4} \left(
			\psi_1^2 - \mu_1^2
	\right)^2 - \frac{\lambda_1}{4} \mu_1^4 + 
	\frac{\lambda_2}{4} \left(
			\psi_2 ^2 - \mu_2^2
		\right)^2 - \frac{\lambda_2}{4} \mu_2^4 + 
		\frac{1}{2} \psi_1^2 \psi_2^2 .
\label{1-132}
\end{eqnarray}
Let us note that the energy functional is defined with accuracy of a constant. We choose the constant as 
$- \frac{\lambda_1}{4} \mu_1^4 - \frac{\lambda_2}{4} \mu_2^4$. In order to understand the physical sense of the obtained solution let us to introduce the energy densities for both BEC's 
\begin{eqnarray}
	\varepsilon_1 &=& 
	\frac{1}{2} \left ( \nabla \psi_1 \right )^2 + 
	\frac{\lambda_1}{4} \left(
			\psi_1^2 - \mu_1^2
	\right)^2 - \frac{\lambda_1}{4} \mu_1^4
\label{1-134}\\
	\varepsilon_2 &=& 
	\frac{1}{2} \left ( \nabla \psi_2 \right )^2 + 
	\frac{\lambda_2}{4} \left(
			\psi_2^2 - \mu_2^2
	\right)^2 - \frac{\lambda_2}{4} \mu_2^4  .
\label{1-138}
\end{eqnarray}
Taking into account the asymptotic behavior \eqref{1-110} \eqref{1-120} we see that at the infinity 
\begin{equation}
	\varepsilon(\psi_{1,2}) \approx - \frac{\lambda_1}{4} \mu_1^4
\label{1-140}
\end{equation}
i.e. is non-zero. It means that we have the space filled with BEC describing by $\psi_1$ and at the center there is a defect filled with BEC describing by $\psi_2$. In Fig. \ref{sec1fig2} the profiles of the energy densities for both BEC's with $\psi_1$ and $\psi_2$ are presented. 

We see that choosing the external potentials $V_i$ with some special values 
\begin{equation}
	V_i = -\lambda_i \frac{\hbar^2 \psi_1^{2/3}(0)}{2 m_1} \mu_i^2 
\label{1-100}
\end{equation}
we have the solution describing a defect filled with one kind of BEC on the background of the space filled with another kind of BEC. From Fig. \ref{sec1fig2} we see the BEC filling the defect displaces the second kind of BEC. Formally it is analogously to Meissner effect in superconductivity. 

\begin{figure}[h]
  \begin{center}
  \fbox{
		\includegraphics[width=0.5\textwidth]{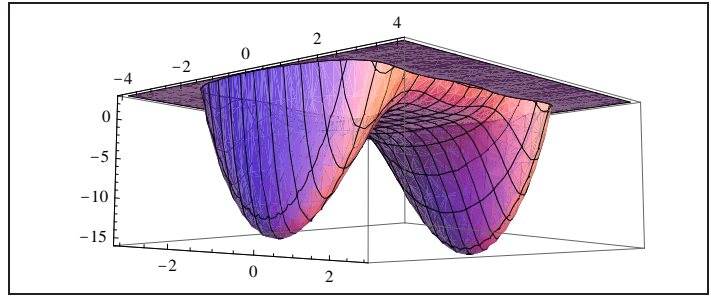}}
	  \caption{The profile of dimensionless potential \eqref{1-130} 
	  as the function of $\psi_{1,2}$. One can see two local and two global minima.}
  \label{sec1fig3}
  \end{center}
\end{figure}

The profile for the dimensionless potential $\varepsilon(\psi_{1,2})$  as the function of functions $\psi_{1,2}$ in Fig.\ref{sec1fig3} is presented. From this figure we see that it has two local and two global minima that gives rise to the regular solutions of Eq's \eqref{1-50} \eqref{1-60}. 

\section{Droplet solution with three and two-body scattering terms and constant external trapping potential}
\label{ThreeTwoScattering}

In this section we consider a droplet solution but BEC's have positive three-body and negative two-body scattering terms on the RHS of Eq's \eqref{2-10} and \eqref{2-20} \cite{Bulgac} . 

In this case a two-component BEC can be described by the two-coupled equations (similar to GP) of the following form 
\begin{eqnarray}
	i \hbar \frac{\partial \tilde \psi_1}{\partial t} &=& 
	\left(
		- \frac{\hbar^2}{2m_1} \nabla^2 + 
		U_{11} \left| \tilde \psi_1 \right|^4 - 
		V_{11} \left| \tilde \psi_1 \right|^2 + 
		U_{12} \left| \tilde \psi_2 \right|^2 + 
		W_1 
	\right) \tilde \psi_1 , 
\label{2-10} \\
	i \hbar \frac{\partial \tilde \psi_2}{\partial t} &=& 
	\left(
		- \frac{\hbar^2}{2m_2} \nabla^2 + 
		U_{22} \left| \tilde \psi_2 \right|^4 - 
		V_{22} \left| \tilde \psi_2 \right|^2 + 
		U_{21} \left| \tilde \psi_1 \right|^2 + 
		W_2 
	\right) \tilde \psi_2 
\label{2-20}
\end{eqnarray}
where  the  notations for $N_i$ are the same as in Section \ref{twoBody}. The resulting equations for the wave functions $\psi_{1,2}(\vec r,t)$ in dimensionless form can be written as
\begin{eqnarray}
	i \frac{\partial \psi_1}{\partial t} &=& 
	\left(
		- \nabla^2 + u_{11} \left| \psi_1 \right|^4 - 
		v_{11} \left| \psi_1 \right|^2 + 
		u_{12} \left| \psi_2 \right|^2 + 
		w_1 
	\right) \psi_1 , 
\label{2-30} \\
	i \frac{\partial \psi_2}{\partial t} &=& 
	\left(
		- k \nabla^2 + u_{22} \left| \psi_2 \right|^4 - 
		v_{22} \left| \psi_2 \right|^2 + 
		u_{21} \left| \psi_1 \right|^2 + 
		w_2 
	\right) \psi_2 
\label{2-40}
\end{eqnarray}
here $k=m_1/m_2$ and we redefined $t/t_0 \rightarrow t$; $\vec r/l_0 \rightarrow \vec r$; $t_0=\frac{2m_1}{\hbar \psi_1^{2/3}(0)}$; $l_0 = \psi_1^{-1/3}(0)$; $\psi_{1,2}=\frac{\tilde \psi_{1,2}}{\psi_1(0)}$; $u_{ii}=\frac{2m_1U_{ii}\psi_1^{10/3}(0)}{\hbar^2}$; $v_{ii}=\frac{2m_1V_{ii}\psi_1^{4/3}(0)}{\hbar^2}$;  $u_{12}=\frac{2m_1U_{12}\psi_1^{4/3}(0)}{\hbar^2}$ and 
$w_{1,2} = \frac{2 m_1 W_{1,2}}{\hbar^2 \psi_1^{2/3}(0)}$. For the simplicity we will  consider the case $k=1$. We are searching for a static  spherical symmetric solution: $\psi_{1,2}(\vec r,t) = \psi_{1,2}(r)$. In this case Eq's \eqref{2-30} \eqref{2-40} are 
\begin{eqnarray}
	\psi_1'' + \frac{2}{r} \psi_1' &=& \psi_1 \left[
		\lambda_3 \psi_2^2 + \lambda_1 \left(  
			\psi_1^2 - \mu_1^2
		\right) \left(  
			3 \psi_1^2 - \mu_1^2
		\right)
	\right],
\label{2-50} \\
	\psi_2'' + \frac{2}{r} \psi_2' &=& \psi_2 \left[
		\lambda_3 \psi_1^2 + 3 \lambda_2 \psi_2^2 \left(  
			\psi_2^2 - \frac{2}{3} \mu_2^2
		\right)
	\right].
\label{2-60}
\end{eqnarray}
where $3 \lambda_{i} = u_{ii}$; $v_1 = 4 \mu_1^2$; $u_{12} = \lambda_3$; $2 \lambda_2 \mu_2^2 = v_{21}$; $\lambda_1 \mu_1^4 = w_1$; $w_2=0$. Eq's \eqref{2-50} \eqref{2-60} are written in the form which is convenient for the numerical calculations. For the numerical calculations we choose the boundary conditions and parameters values as \eqref{1-70} - \eqref{1-90}. 

\begin{figure}[h]
\begin{minipage}[t]{.45\linewidth}
  \begin{center}
  \fbox{
  \includegraphics[width=1.\textwidth]{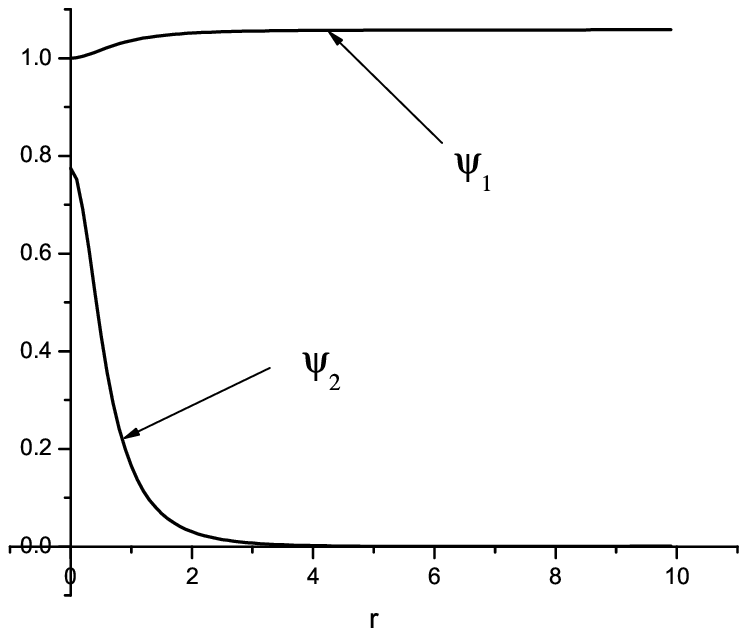}}
  \caption{The profiles of $\psi_{1,2}(x)$, $\mu_1 = 1.0576294$, 
  $\mu_2 = 4.05682$.}
  \label{sec3_fig1}
  \end{center}
\end{minipage}\hfill
\begin{minipage}[t]{.45\linewidth}
  \begin{center}
  \fbox{
		\includegraphics[width=1.\textwidth]{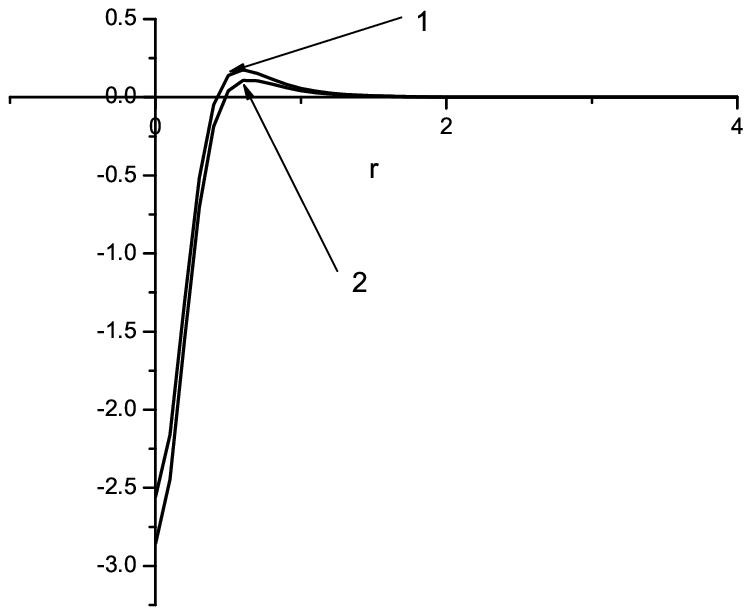}}
	  \caption{The profiles of the dimensionless energy densities: 
	  the curve 1 - the dimensionless energy density \eqref{2-100}
	  for both BEC's;
	  the curve 2 - the dimensionless energy density \eqref{2-130}
	  for $\psi_2$ BEC.}
  \label{sec3_fig2}
  \end{center}
\end{minipage}\hfill 
\end{figure}

Again we search the solution as the nonlinear eigenvalue problem: $\mu_{1,2}$ are eigenvalues and corresponding functions $\psi_{1,2}$ are eigenfunctions. The profiles of functions $\psi_{1,2}$ in Fig. \ref{sec3_fig1} are presented. 

\begin{figure}[h]
\begin{minipage}[t]{.45\linewidth}
  \begin{center}
  \fbox{
		\includegraphics[width=1.\textwidth]{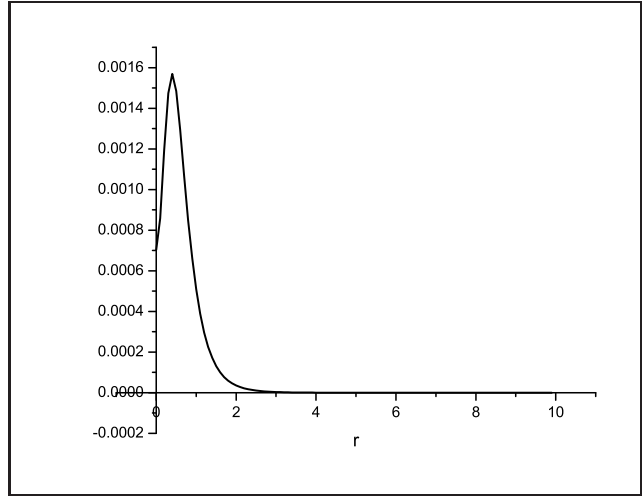}}
	  \caption{The profile of the dimensionless energy density 
	  \eqref{2-110} for $\psi_1$ BEC.}
  \label{sec3_fig3}
  \end{center}
\end{minipage}\hfill 
\end{figure}

The asymptotic behavior of the functions $\psi_{1,2}$ for such two  interacting BEC's is following 
\begin{eqnarray}
	\psi_1 & \approx & \mu_1 - \psi_{1, \infty} 
	\frac{e^{-r \sqrt{4 \lambda_1 \mu_1^4}}}{r} , 
\label{2-70} \\
	\psi_2 & \approx & \psi_{2, \infty} 
	\frac{e^{-\mu_1 r}}{r}
\label{2-80} 
\end{eqnarray}
where $\psi_{1,2, \infty}$ are constants. 

One can consider Eq's \eqref{2-50} \eqref{2-60} as Euler - Lagrangian equations. Then the energy density has the form 
\begin{eqnarray}
	\varepsilon \left( \psi_{1,2} \right) &=&  
	\frac{1}{2} \left ( \nabla \psi_1 \right )^2 + 
	\frac{1}{2} \left ( \nabla \psi_2 \right )^2 + 
		V \left( \psi_{1,2} \right) ,
\label{2-100}\\
	V \left( \psi_{1,2} \right) &=& 
		\frac{\lambda_1}{2} \psi_1^2 \left( 
			 \psi_1^2 - m_1^2 
		\right)^2 + 
		\frac{\lambda_2}{2} \psi_2^4 \left( 
			 \psi_2^2 - m_2^2 
		\right) + 
		\frac{1}{2} \psi_1^2 \psi_2^2 .
\label{2-105}
\end{eqnarray}
The energy densities for both BEC's are 
\begin{eqnarray}
	\varepsilon_1 &=& 
	\frac{1}{2} \left ( \nabla \psi_1 \right )^2 + 
	\frac{\lambda_1}{2} \psi_1^2 \left(
			\psi_1^2 - \mu_1^2
	\right)^2  
\label{2-110}\\
	\varepsilon_2 &=& 
	\frac{1}{2} \left ( \nabla \psi_2 \right )^2 + 
	\frac{\lambda_2}{2} \psi_2^4 \left(
			\psi_2^2 - \mu_2^2
	\right)  .
\label{2-130}
\end{eqnarray}
Taking into account the asymptotic behavior \eqref{2-70} \eqref{2-80} we see that at the infinity 
\begin{equation}
	\varepsilon(\psi_{1,2}) \approx 0 .
\label{2-140}
\end{equation}
It means that we have a droplet filled with two interacting BEC's and trapped with the external potentials 
\begin{equation}
	w_1 = \lambda_1 \mu_1^4 \text{ and }
	w_2 = 0. 
\label{2-150}
\end{equation}
In Fig's \ref{sec3_fig2} \ref{sec3_fig3} the profiles of the energy densities for both BEC's and for $\psi_1$ BEC and for $\psi_2$ BEC are presented. 

\section{Droplet solution with four and three-body scattering terms and without external trapping potential}
\label{dropletWithout}

In this section we would like to consider the interaction between two   BEC's where one BEC has a \emph{hypothesized positive strong four-body scattering term}. The first BEC has three and two-body scattering (analogously to Section \ref{ThreeTwoScattering}) but the second BEC has four and three-body scattering terms. In order to obtain a \emph{self-maintaining droplet} without any external trapping potential we assume that there exists BEC with positive four-body scattering and negative three-body scattering. 

In this case BEC's equations are 
\begin{eqnarray}
	i \hbar \frac{\partial \tilde \psi_1}{\partial t} &=& 
	\left(
		- \frac{\hbar^2}{2m_1} \nabla^2 + 
		W_{11} \left| \tilde \psi_1 \right|^6 - 
		U_{11} \left| \tilde \psi_1 \right|^4 + 
		V_{11} \left| \tilde \psi_1 \right|^2 + 
		U_{12} \left| \tilde \psi_2 \right|^2 
	\right) \tilde \psi_1 , 
\label{3-10} \\
	i \hbar \frac{\partial \tilde \psi_2}{\partial t} &=& 
	\left(
		- \frac{\hbar^2}{2m_2} \nabla^2 + 
		U_{22} \left| \tilde \psi_2 \right|^4 - 
		V_{22} \left| \tilde \psi_2 \right|^2 + 
		U_{21} \left| \tilde \psi_1 \right|^2 
	\right) \tilde \psi_2 
\label{3-20}
\end{eqnarray}
The resulting equations for the wave functions $\psi_{1,2}(\vec r,t)$ in dimensionless form can be written as
\begin{eqnarray}
	i \frac{\partial \psi_1}{\partial t} &=& 
	\left(
		- \nabla^2 + w_{11} \left| \psi_1 \right|^6 - 
		u_{11} \left| \psi_1 \right|^4 + 
		v_{11} \left| \psi_1 \right|^2 + 
		u_{12} \left| \psi_2 \right|^2 
	\right) \psi_1 , 
\label{3-30} \\
	i \frac{\partial \psi_2}{\partial t} &=& 
	\left(
		- k \nabla^2 + 
		u_{22} \left| \psi_2 \right|^4 - 
		v_{22} \left| \psi_2 \right|^2 + 
		u_{21} \left| \psi_1 \right|^2 
	\right) \psi_2 
\label{3-40}
\end{eqnarray}
here $k=m_1/m_2$ and we redefined 
$t/t_0 \rightarrow t$; $\vec r/l_0 \rightarrow \vec r$; 
$t_0=\frac{2m_1}{\hbar \psi_1^{2/3}(0)}$; 
$l_0 = \psi_1^{-1/3}(0)$; 
$\psi_{1,2}=\frac{\tilde \psi_{1,2}}{\psi_1(0)}$; 
$w_{11}=\frac{2m_1 \psi_1^{16/3}(0)}{\hbar^2} W_{11}$;  
$u_{ii} = \frac{2m_1 \psi_1^{10/3}(0)}{\hbar^2} U_{ii}$; 
$v_{ii} = \frac{2m_1 \psi_1^{4/3}(0)}{\hbar^2} V_{ii}$ and 
$u_{ij}=\frac{2m_1 \psi_1^{4/3}(0)}{\hbar^2} U_{ij}$. For the simplicity we will  consider the case $k=1$. We are searching for a static  spherical symmetric solution: $\psi_{1,2}(\vec r,t) = \psi_{1,2}(r)$. In this case Eq's \eqref{3-30} \eqref{3-40} are 
\begin{eqnarray}
	\psi_1'' + \frac{2}{r} \psi_1' &=& \psi_1 \left[
		\lambda_3 \psi_2^2 + \lambda_1 \psi_1^2 \left(  
			\psi_1^2 - \mu_1^2
		\right) \left(  
			2 \psi_1^2 - \mu_1^2
		\right)
	\right],
\label{3-50} \\
	\psi_2'' + \frac{2}{r} \psi_2' &=& \psi_2 \left[
		\lambda_3 \psi_1^2 + 3 \lambda_2 \psi_2^2 \left(  
			\psi_2^2 - \frac{2}{3} \mu_2^2
		\right)
	\right].
\label{3-60}
\end{eqnarray}
where $2 \lambda_{1} = w_{11}$; 
$3 \lambda_{1} \mu_{i}^2 = u_{11}$; 
$\lambda_1 \mu_1^4 = v_{11}$;
$\lambda_3 = u_{12} = u_{21}$; 
$3 \lambda_2 = u_{22}$ and 
$2 \lambda_2 \mu_1^2 = v_{22}$. Eq's \eqref{3-50} \eqref{3-60} are written in the form which is convenient for the numerical calculations. For the numerical calculations we choose the boundary conditions and parameters values as in \eqref{1-70}-\eqref{1-90}. 

The solution is searching as the nonlinear eigenvalue problem: $\mu_{1,2}$ are eigenvalues and corresponding functions $\psi_{1,2}$ are eigenfunctions. We solve equations set \eqref{3-50} \eqref{3-60} numerically. The profiles of functions $\psi_{1,2}$ practically coincides with the profiles of the functions  $\psi_{1,2}$ from Section \ref{ThreeTwoScattering}. 

The asymptotic behavior of the functions $\psi_{1,2}$ is following 
\begin{eqnarray}
	\psi_1 & \approx & \mu_1 - \psi_{1, \infty} 
	\frac{e^{-r \sqrt{2 \lambda_1 \mu_1^6}}}{r} , 
\label{3-70} \\
	\psi_2 & \approx & \psi_{2, \infty} 
	\frac{e^{-\mu_1 r}}{r}
\label{3-80} 
\end{eqnarray}
where $\psi_{1,2, \infty}$ are constants. 

One can consider Eq's \eqref{3-50} \eqref{3-60} as Euler - Lagrangian equations. Then the dimensionless energy density has the form 
\begin{eqnarray}
	\varepsilon(\psi_{1,2}) &=& 
	\frac{1}{2} \left ( \nabla \psi_1 \right )^2 + 
	\frac{1}{2} \left ( \nabla \psi_2 \right )^2 + 
	V \left( \psi_{1,2} \right) ,
\label{3-90}\\
	V \left( \psi_{1,2} \right) &=& 
	\frac{\lambda_1}{4} \psi_1^4 \left(
			\psi_1^2 - \mu_1^2
	\right)^2  + \frac{\lambda_2}{2} \psi_2 ^4 \left(
			\psi_2^2 - \mu_2^2
		\right) + \frac{1}{2} \psi_1^2 
		\psi_2^2 .
\label{3-95}
\end{eqnarray}
The energy densities for both BEC's are 
\begin{eqnarray}
	\varepsilon_1 &=& 
	\frac{1}{2} \left ( \nabla \psi_1 \right )^2 + 
	\frac{\lambda_1}{4} \psi_1^4 \left(
			\psi_1^2 - \mu_1^2
	\right)^2 , 
\label{3-100}\\
	\varepsilon_2 &=& 
	\frac{1}{2} \left ( \nabla \psi_2 \right )^2 + 
	\frac{\lambda_2}{2} \psi_2^4 \left(
			\psi_2^2 - \mu_2^2
		\right)  .
\label{3-110}
\end{eqnarray}
Taking into account the asymptotic behavior \eqref{3-70} \eqref{3-80} we see that at the infinity 
\begin{equation}
	\varepsilon(\psi_{1,2}) \approx 0 .
\label{3-120}
\end{equation}
It means that we have obtained a self-maintaining droplet filled with two BEC. The profiles of the dimensionless energy densities \eqref{3-90}, \eqref{3-100} and \eqref{3-110} practically do not differ from the corresponding functions from Section \ref{ThreeTwoScattering}.

\section{Discussion and conclusions}

From the mathematical point of view \emph{regular} solutions (having either a finite energy or at least finite values of the fields) for two interacting BEC's do exist only because the corresponding potential either \eqref{1-132} or \eqref{2-105} or \eqref{3-95} has local and global minima. If this potential has only a global minimum then following Derrick's theorem  \cite{derrick} such solution may exist in the dimension 1+1 only \footnote{under the condition of absence other fields: for instance, t'Hooft - Polyakov monopole solution \cite{thooft} is regular one and has only a global minimum but there exist additional fields -- SU(2) Yang-Mills gauge fields} and this solution is known as a kink \footnote{it should be reminded that Derrick in his paper specifically states that his theorem is altered if there are multiple fields or higher derivatives involved}. It is interesting to compare the properties of presented here solution with the kink solution. The main difference is that the solutions obtained here is topological trivial whereas the kink is topological non-trivial solution. The topological non-triviality means that there exists an integer (a topological charge) that does not change by a small deviation of the solution. Usually the topological charge is characterized by the solution behavior at the infinity. Let us emphasize once again that the solutions obtained here are topological trivial and they exist because the corresponding potential has both local and global minima only. 

Now we would like to discuss the physical interpretation of the obtained solutions. The interpretation strongly depends on the asymptotic value of the energy densities. In the case of non-zero value we have the space filled with one Bose - Einstein condensate $\psi_1$ and on the background of this condensate there is a spherically symmetric defect (cavity) filled with another Bose - Einstein condensate $\psi_2$. The asymptotic behavior of both condensates is different: for $\psi_1$ condensate the asymptotic value is non-zero but for $\psi_2$ condensate the asymptotic value is zero. The same is valid for both energy densities. If the asymptotic value is zero we have droplet solution: there exists a ball filled with two  BEC's. There are two possibilities: either the droplet is trapped with an \emph{constant} external potential (Section \ref{ThreeTwoScattering}) or not trapped without any external potential (Section \ref{dropletWithout}). 

Numerical calculations show that most likely the existence of a regular solution depends on the potential form only: whether has it or not local and global minima simultaneously. In this connection one can mention that similar solution have been found for a scalar model of a glueball \cite{Dzhunushaliev:2011we}. In this model the Lagrangian from SU(3) gauge theory by some manner can be approximated by a Lagrangian with two scalar fields. These two scalar fields approximately describe 2-$th$ and 4-$th$ Green functions (i.e. the correlation between fields in two or four points in the spacetime) of SU(3) gauge fields. The first scalar field describes a gauge field belonging to subgroup $SU(2) \subset SU(3)$ and the second scalar field describes fields belonging to a coset $SU(3) / SU(2)$. The kinetic terms from the initial Lagrangian gives rise to kinetic terms for both scalar fields. The terms like $A^B_\mu A^C_\nu A^D_\rho A^E_\sigma$ give rise to terms like $(\psi_i^2 - \psi_{i; \infty}^2)^2$ in corresponding potentials for the scalar fields $\psi_i$ $(i=1,2)$. Here $A^B_\mu$ is the SU(3) gauge potential. The difference between the potential \eqref{2-100} and the potential obtained in Ref. \cite{Dzhunushaliev:2011we} is that the potential \eqref{2-100} is the polynomial of 6-$th$ order whereas the potential from \cite{Dzhunushaliev:2011we} is polynomial of 4-$th$ order. But in both cases corresponding potentials have local and global minima that leads to the existence of a regular solution. In Ref. \cite{Bulgac} a similar construction is considered and following conclusion is made "\ldots a dilute Fermi - Dirac droplet will behave very much like a nuclear system."

Finishing the comparison of obtained here solution with the solution obtained for the scalar model of a glueball we see that there is clear analogy between both solutions. It allows us to draw interesting and useful analogy between physical objects from high energy physics and BEC's physics: (a) the defect solution is similar to a cavity on the background of the space filled with a nonzero gluon condensate; (b) the droplet solution is similar to a glueball. 

Another important feature of the solution obtained here is following: In Ref. \cite{Cirac} it was shown that under certain conditions, the system of two BEC's has an almost degenerate ground state which is separated from the excited levels by an energy gap. The atomic ensemble then behaves like a two-level system that could be used to encode a qubit. The authors have offered the idea of using these two many-body states to encode a qubit and use it for quantum computation. One can use the defect/droplet obtained here as an elementary element (quantum gate) for creating the qubit device offered in Ref. \cite{Cirac}. 

Finally we would like to list the main features of the solutions obtained here:
\begin{itemize}
\item The presented defect/droplet solutions exist for two interacting BEC's only.
\item The solutions exist for special choice of the BEC's potential: it should have local and global minimum only.
\item The solutions are eigenfunctions for nonlinear differential equations. Consequently some physical parameters must have a well defined values (external potential, coefficients in front of $n-$body scattering terms and so on).
\item The droplet solution without any external trapping potential exists in the case of very strong interaction between atoms of BEC only. The droplet solution presented here does exist if there is a positive four-body scattering term of atoms for one BEC. 
\item The defect solution exists for the standard potential of two interacting BEC's but with some special choice external trapping 
potentials. 
\end{itemize}

\section*{Acknowledgments}

I am grateful to the Research Group Linkage Programme of the Alexander von
Humboldt Foundation for the support of this research. Special thanks for A. Avdeenkov for the fruitful discussion.


\begin{thebibliography}{99}

\bibitem{Ho}
T.L. Ho, V.B. Shenoy, Phys. Rev. Lett. 77 (1996) 3276;\\
B.D. Esry, C.H. Greene, J.P. Burke, J.L. Bohn, Phys. Rev. Lett. 78 (1997) 3594;\\
H. Pu, N.P. Bigelow, Phys. Rev. Lett. 80 (1998) 1130.

\bibitem{Bulgac}
Aurel Bulgac, 
%``Dilute Quantum Droplets'', 
Phys. Rev. Lett. 89, 050402 (2002).

\bibitem{Kumar}
V. Ramesh Kumar, R. Radha, Miki Wadati, 
%``Collision of bright vector solitons in two-component BoseEinstein condensates'', 
Phys. Lett. A, 374 (2010) 36853694.

\bibitem{QiuYan}
Li Qiu-Yan, Li Zai-Dong, Yao Shu-Fang), Li Lu, and Fu Guang-Sheng, 
%``Formation of combined solitons in two-component Bose  Einstein condensates'', 
Chin. Phys. B, Vol.19, No. 8 (2010) 080501. 

\bibitem{Deng}
Deng-Shan Wang, Xing-Hua Hu and W. M. Liu, 
%``Localized nonlinear matter waves in two-component Bose-Einstein condensates with time- and space-modulated nonlinearities'', 
Phys. Rev. A, 82, 023612 (2010). 

\bibitem{Baizakov}
B. B. Baizakov, A M Kamchatnov and M Salerno, 
%``Matter sound waves in two-component BoseEinstein condensates'', 
J. Phys. B: At. Mol. Opt. Phys. 41 (2008) 215302.

\bibitem{Kawaja}
Al Kawaja and Stoof, 
Nature 411, 918 (2001);\\
J.~Ruostekoski and J.~R.~Anglin,
  %``Creating vortex rings and three-dimensional skyrmions in Bose-Einstein condensates,''
  Phys.\ Rev.\ Lett.\  {\bf 86}, 3934 (2001)
  [cond-mat/0103310];\\
C.~M.~Savage and J.~Ruostekoski,
  %``Energetically stable particle-like skyrmions in a trapped Bose-Einstein condensate,''
  Phys.\ Rev.\ Lett.\  {\bf 91}, 010403 (2003)
  [cond-mat/0306112]; \\
J.~Ruostekoski,
  %``Stable particlelike solitons with multiply-quantized vortex lines in Bose-Einstein condensates,''
  Phys.\ Rev.\ A {\bf 70}, 041601 (2004)
  [cond-mat/0408376].
  
\bibitem{Law}
K. J. H. Law, P. G. Kevrekidis and Laurette S. Tuckerman, 
``Stable Vortex-Bright Soliton Structures in Two-Component Bose Einstein Condensates', 
arXiv:1001.4835. 

\bibitem{Zhang}
Xiao-Fei Zhang, Xing-Hua Hu, Xun-Xu Liu and W. M. Liu, 
%``Vector solitons in two-component Bose-Einstein condensates with tunable interactions and harmonic potential'', 
Phys. Rev. A 79, 033630 (2009). 

\bibitem{derrick} 
G.H. Derrick, J. Math Phys. {\bf 5} 1252 (1964).

\bibitem{thooft} 
G. 't Hooft, Nucl. Phys. B {\bf 79}, 276 (1974); \\
A.M. Polyakov, JETP {\bf 41}, 988 (1975).

\bibitem{Dzhunushaliev:2011we}
  V.~Dzhunushaliev,
  ``SU(3) glueball gluon condensate,''  
  [arXiv:1110.1427 [hep-ph]].

\bibitem{Cirac}
 J. I. Cirac, M. Lewenstein, K. Molmer, and P. Zoller, 
 ``Quantum superposition states of Bose-Einstein condensates'', 
 Phys. Rev. A, 57(2):12081218, 1998.

\end{thebibliography}
\end{document}